\documentclass[conference]{IEEEtran}

\usepackage{amsmath,amssymb,graphicx,xcolor,hyperref,fancyhdr,bm,subfigure}
\definecolor{metablue}{RGB}{0,141,169}

\hypersetup{
	colorlinks=true,
	linkcolor=metablue,
	citecolor=metablue,
	urlcolor=metablue
}

\makeatletter
\fancypagestyle{plain}{
	\fancyhf{}
	\fancyhead[C]{META 2026, DUBLIN -- IRELAND, JULY 14 -- 17, 2026}
	
}
\fancypagestyle{IEEEtitlepagestyle}{
	\fancyhf{}
	\fancyhead[C]{META 2026, DUBLIN -- IRELAND, JULY 14 -- 17, 2026}
	
}
\makeatother

\title{Frequency Conversion Characteristics of Spatiotemporal Josephson Metasurfaces for Quantum Applications}

\author{
	Sajjad Taravati\\[1ex]
	Faculty of Engineering and Physical Sciences, University of Southampton, Southampton SO17 1BJ, UK \\
	Corresponding author: \href{mailto:s.taravati@soton.ac.uk}{s.taravati@soton.ac.uk}
}

\begin{document}
	\maketitle
	
	\begin{abstract}
This presentation explores the various characteristics of a nonreciprocal, frequency-converting Josephson metasurface operating at millikelvin temperatures. Leveraging the unique properties of Josephson junctions, which support supercurrent flow without resistance, this metasurface enables efficient manipulation of nonlinear wave interactions, facilitating both frequency conversion and amplification of incident photons. 
\end{abstract}

Temporal modulation has garnered significant attention due to its wide-ranging applications in modern wireless communication systems, photonics, radar technologies, and beyond~\cite{Taravati_ACSP_2022,taravati20234d,Taravati_NC_2021}. It exhibits dynamic properties characterized by the modulation of electrical permittivity, magnetic permeability, or electrical conductivity across both space and time~\cite{Taravati_Kishk_MicMag_2019,Taravati_PRAp_2018,Taravati_Kishk_PRB_2018,Taravati_Kishk_TAP_2019,taravati2020full,taravati2024finite}. Understanding their behavior is crucial for designing advanced devices and systems with enhanced functionality and performance. Space-time metamaterials can function as perfect isolator through one-way time coherency~\cite{taravati2017self}, perfect absorbers enabled by time coherency~\cite{taravati2024spatiotemporal}, unidirectional beam splitter and amplifier through time coherency~\cite{Taravati_Kishk_PRB_2018}, four-dimensional surface-wave generator~\cite{taravati2019_mix_ant,Taravati_AMA_PRApp_2020}, nonreciprocal transmission platforms~\cite{taravati2020full,taravati2021nonreciprocal,taravati2024spatiotemporal}, spatial isolators~\cite{lira2012electrically,Taravati_PRB_SB_2017,Taravati_AMTech_2021}, frequency convertors~\cite{taravati2021pure,Taravati_PRB_Mixer_2018}, circulators~\cite{taravati2022low}, amplifiers~\cite{taravati2025temporal}, multiple access secure communication systems~\cite{taravati_PRApp_2019}, and multifunctional antenna systems~\cite{Taravati_AMA_PRApp_2020}. However, traditional time-modulated devices and frequency converters and associated electronic components, such as varactors, transistors, and diodes, fall short in the millikelvin-temperature environments of superconducting quantum technologies due to their operational limitations and the noise they introduce. 

Here, we present a quantum-compatible nonreciprocal frequency converter Josephson metasurface operating at millikelvin temperatures~\cite{taravati2024nonlinear,taravati2024one,taravati2025light,taravati2025_entangle,taravati2025frequency}, leverage the quantum mechanical phenomenon of supercurrent flow without resistance, offering unparalleled opportunities for nonlinear wave manipulation. Key features of our approach include efficient frequency conversion of incident photons, along with significant frequency conversion and amplification. In linear STM media, the wave equation produces coupled differential equations for each harmonic, with low modulation frequencies allowing strong coupling and efficient sideband generation. Nonlinear systems, however, introduce higher-order terms that enable efficient harmonic generation even at high modulation frequencies. Space-time-modulated Josephson metasurfaces achieve high-efficiency frequency conversion at large modulation-to-signal frequency ratios. Unlike linear systems, which are constrained by phase-matching and dispersion and require long interaction lengths, spatiotemporal Josephson metasurface  deliver superior performance in a compact size, making them ideal for millikelvin temperature superconducting quantum technologies.

\begin{figure}
		\begin{center}
			\includegraphics[width=1\columnwidth]{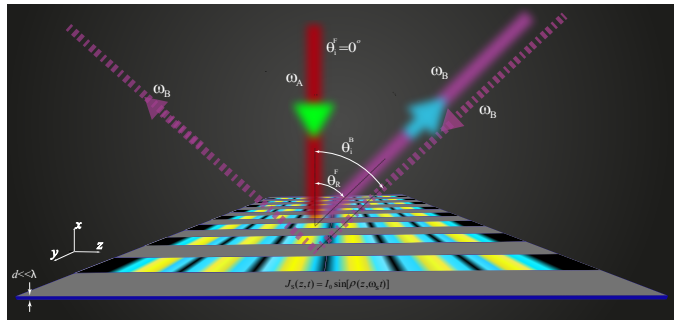}
			\caption{Nonreciprocal frequency conversion of spatiotemporal Josephson metasurface.}
			\label{Fig:sch}
		\end{center}
\end{figure}

Figure~\ref{Fig:sch} illustrates nonreciprocal frequency conversion a spatiotemporal Josephson metasurface, featuring an array of cascaded space-time-modulated Josephson junctions. The metasurface offers a unique one-way frequency conversion from $\hbar \omega_\text{A}$ to $\hbar \omega_\text{B}$. The metasurface is characterized by a space-time-varying current density, expressed as $J(z,t) = I_0 \sin[\rho(z,t)]$. When a signal wave with frequency $\omega_0$ is normally incident on the metasurface, and the modulation frequency is set to $\omega_\text{s}$, a pure frequency conversion occurs, resulting in a new frequency $\omega_0 + \omega_\text{s}$. The converted signal exhibits conversion gain, demonstrating highly efficient frequency conversion with the desired high/low conversion frequency ratio, conversion gain, and spurious-free operation. This behavior is typical of the proposed metasurface, which may be further tailored to achieve other desired frequency conversion outcomes—such as down-conversion, low/high frequency conversion, or different conversion gains—through appropriate design of the space-time modulation's band structure and parameters. 

Given the periodicity of the metasurface in space and time, the magnetic field in the metasurface may be decomposed to Floquet space-time harmonics, as $\mathbf{H}_\text{s}(x,z,t)=\mathbf{\hat{y}}\sum_{n }   \varPsi_{n}  e^{ - j \left[ k_x x+ \kappa_{n} z -\omega_n t\right]}$. Considering the incident field $\mathbf{H}_\text{I} (x,z,t)= \mathbf{\hat{y}} H_0 e^{j \omega _0 t}\cdot e^{-j\left[k_0\sin(\theta_\text{i}) x +k_0 \cos(\theta_\text{i}) z \right]}$, the reflected fields reads
\begin{equation}
		\mathbf{H}_\text{R} (x,z,t)= \mathbf{\hat{y}} \sum\limits_{n =  - \infty }^\infty  R_{n} e^{-j \left[ k_{0} \sin(\theta_{\text{i}}) x -k_{0n} \cos(\theta_n^\text{R}) z  -\omega_n t\right] } ,
		\label{eqa:A-E_refl_forw}
\end{equation}
\noindent where $\theta_\text{i}$ is defined as the angle between the incident wave and the metasurface boundary, and $\theta_n^\text{R}$ is defined as the angle between the $n$th reflected space-time harmonic and the metasurface boundary. We apply continuity of the tangential components of the electromagnetic fields at $z=0$ and $z=d$ to find the unknown field amplitudes $R_{n}$. 

To demonstrate the capability of the proposed apparatus in nonreciprocal pure frequency conversion, we design a metasurface with the thickness $d=0.1 \lambda$, and the space-time-varying permeability given by 
\begin{equation}
\mu_\text{s}(z,t)= \sec(0.75+ 0.78 \sin[2\pi\times8.5\times 10^{9}( z/c- t)]).
\end{equation}

\begin{figure}
	\begin{center}	
					\vspace*{0.4cm}		
		\subfigure[]{\label{Fig:res1}
			\includegraphics[width=1.05\columnwidth]{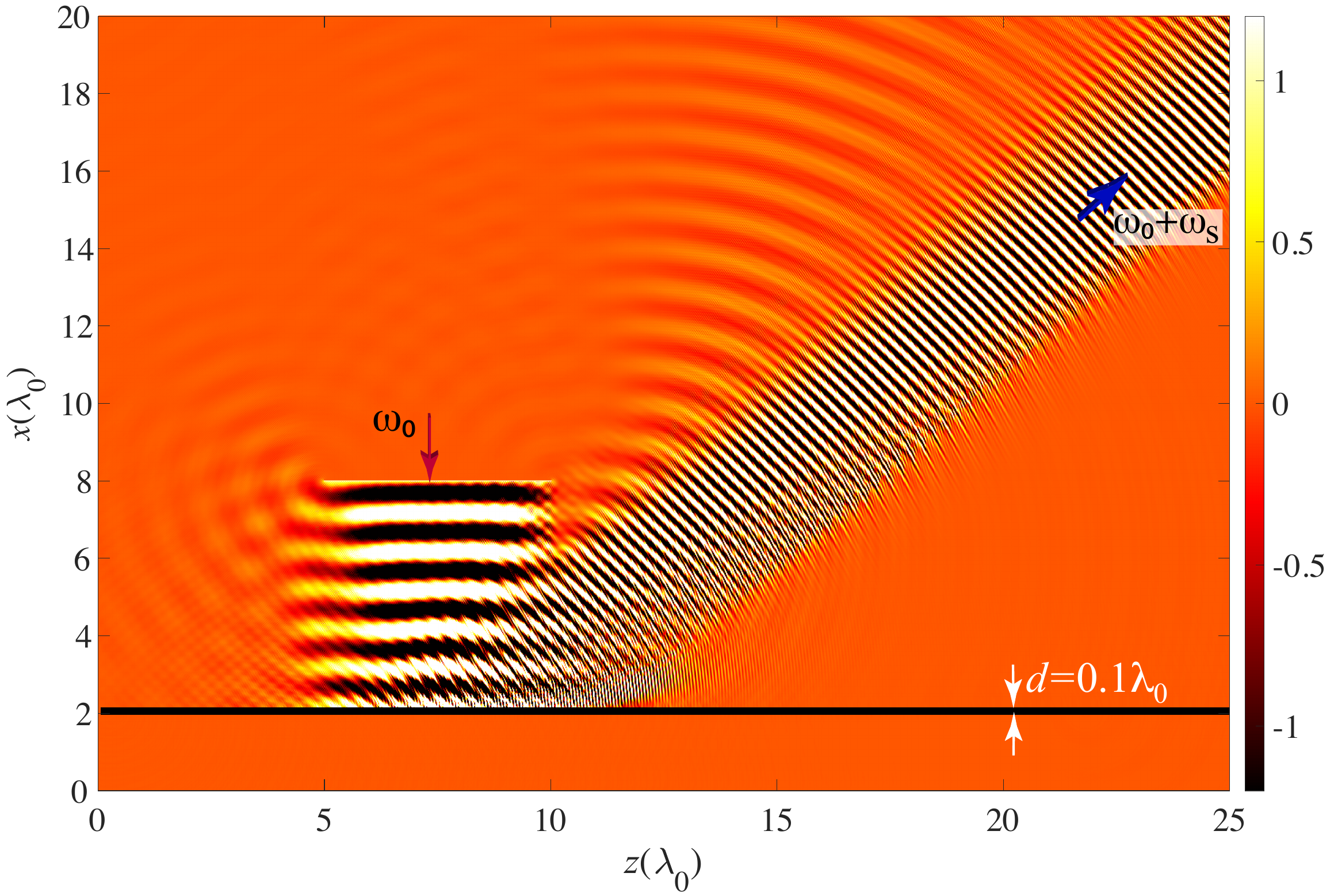}} 
		\hfil
		\subfigure[]{\label{Fig:spec-a} 
			\includegraphics[width=0.48\columnwidth]{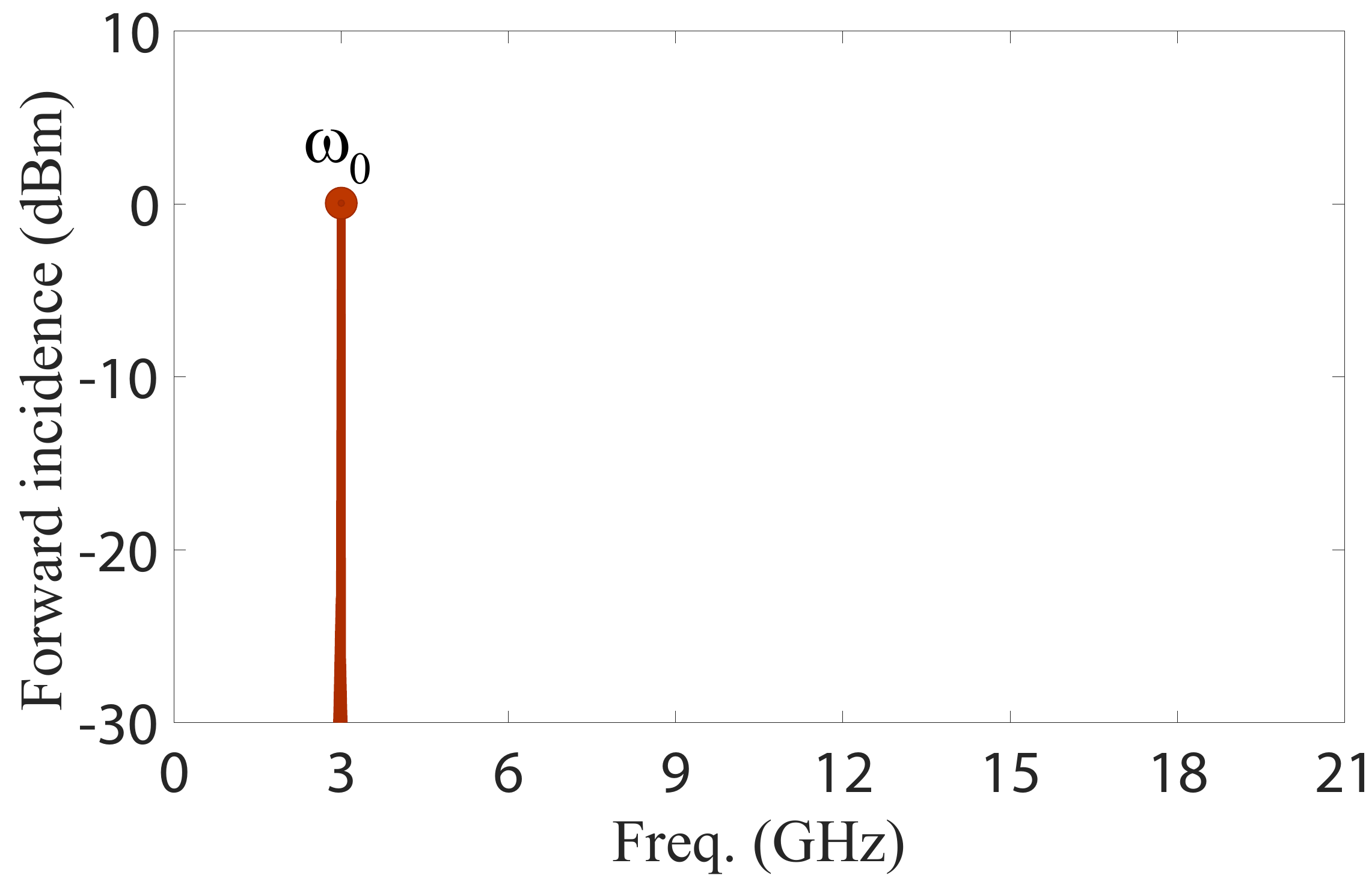}}
					\hfil
			\subfigure[]{\label{Fig:spec-c}
				\includegraphics[width=0.48\columnwidth]{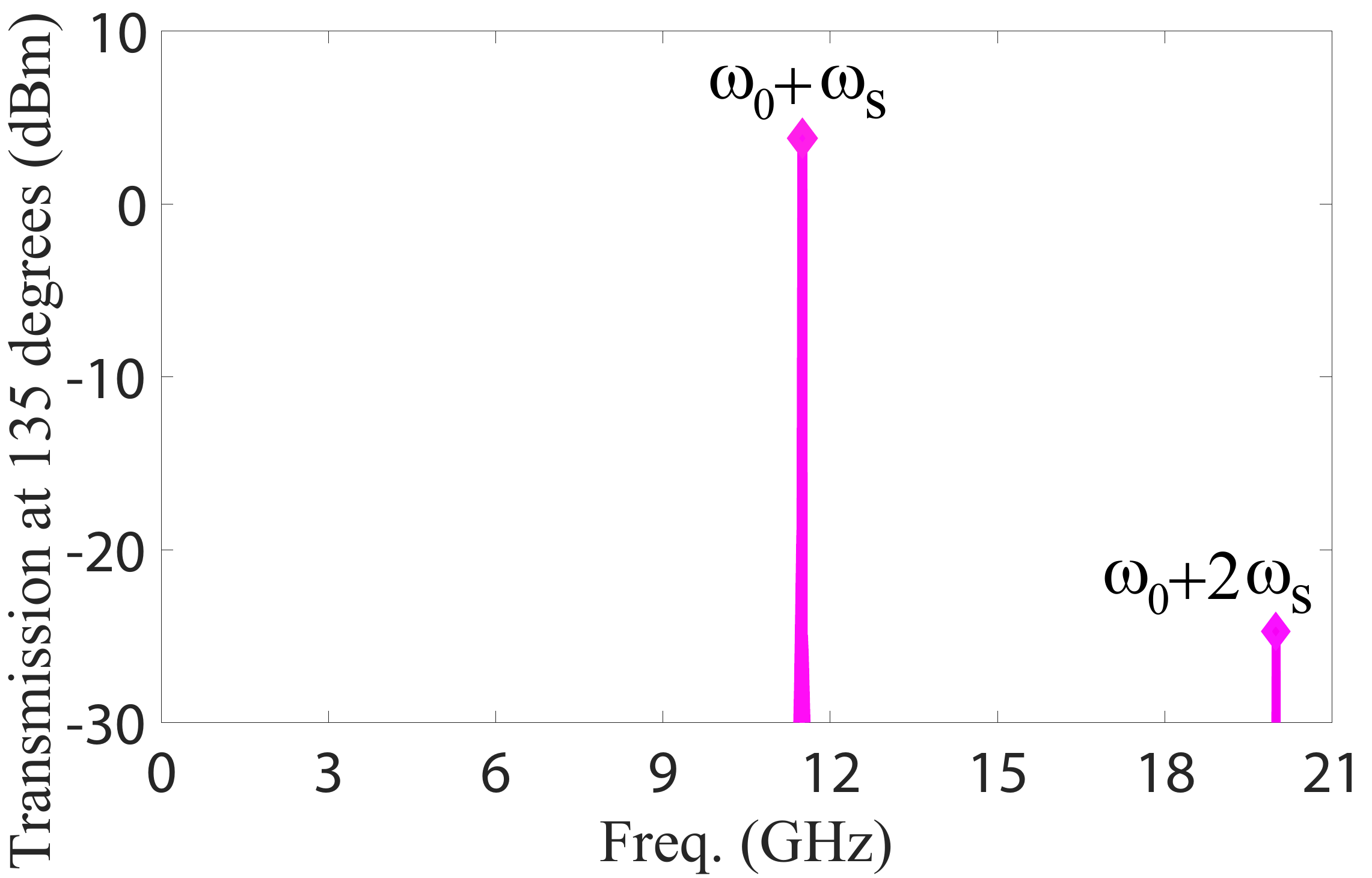}}
		\caption{Pure quantum frequency conversion by a space-time-modulated Josephson junction metasurface. (a)~FDTD numerical simulation results for magnetic field distribution $H_y$. Normal excitation results in a highly efficient frequency conversion from $\omega_0=2\pi \times 3$~GHz to $\omega_0+\omega_\text{s}=2\pi \times 11.5$~GHz. (b) and (c) Frequency spectrum of the incident wave and the \textit{up-converted} reflected wave at $\theta_\text{r}=135^\circ$, respectively.} 
		\label{fig:3}
	\end{center}
\end{figure}

Figure~\ref{Fig:res1} presents the simulation results for forward wave excitation, where a wave at $\omega_0=2\pi \times 3$~GHz is directed at an angle of $\theta=0^\circ$. The incident wave is reflected by the metasurface, which has a thickness of $d=0.1 \lambda$, undergoing a pure frequency up-conversion from $\omega_0=2\pi \times 3$~GHz to $\omega_0+\omega_\text{s}=2\pi \times 11.5$~GHz.

Figure~\ref{Fig:res1} presents the FDTD numerical results for the magnetic field distribution $H_y$, illustrating the frequency conversion capability of the proposed spatiotemporal metasurface. The metasurface, composed of a semiconductor-superconductor heterostructure, is excited normally, resulting in a highly efficient frequency conversion from the initial state at \( \omega_0 = 2\pi \times 3 \) GHz to a higher state at \( \omega_0 + \omega_\text{s} = 2\pi \times 11.5 \) GHz. This process demonstrates the effective manipulation of quantum states associated with different frequency domains. A notable conversion gain of 4.46 dB is achieved, underscoring the high efficiency of the proposed metasurface. The frequency conversion ratio of 3.83 further demonstrates the system's potential for practical quantum applications. This capability highlights the metasurface's role in enabling efficient quantum state transfer across different energy levels, a critical feature for advancing quantum communication and computing technologies. Figures~\ref{Fig:spec-a} and~\ref{Fig:spec-c} plot the the frequency spectrum of the incident wave and the transmitted up-converted wave at $\omega_0+\omega_\text{s}=2\pi \times 11.5$~GHz under a reflection angle of $\theta_\text{r}=135^\circ$, as shown in Fig.~\ref{Fig:res1}. The observed spectral purity in the converted wave highlights the metasurface's remarkable ability to achieve a pristine frequency conversion process, devoid of unwanted mixing products or spurious signals.

\begin{figure}
	\begin{center}
		\includegraphics[width=1\columnwidth]{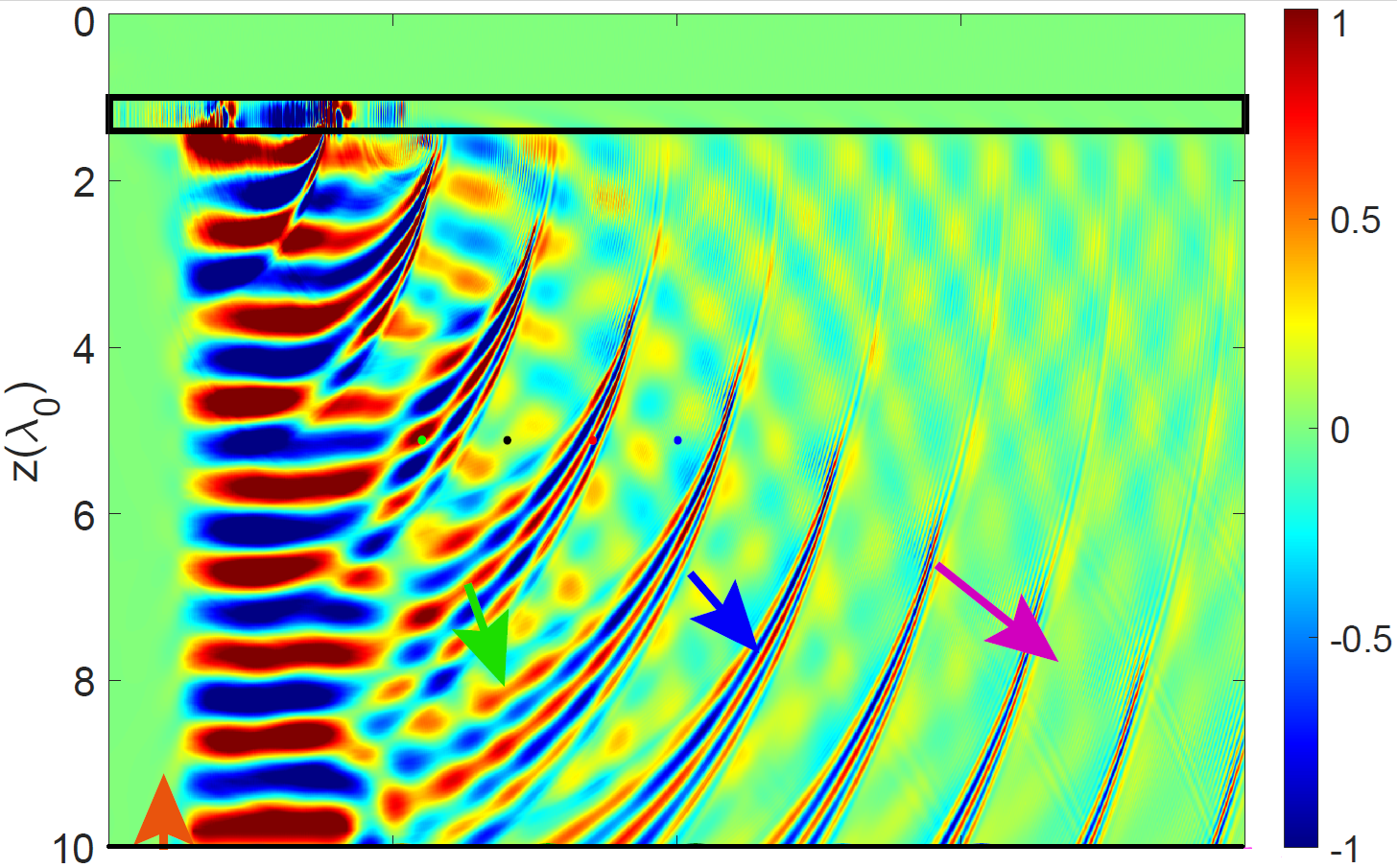}
		\caption{FDTD simulation of a space-time-modulated Josephson metasurface performing frequency mixing at \( f_\text{s} = f_\text{0} = 1.5 \, \text{GHz} \), generating multiple harmonics from \( 2f_\text{s} \) to \( 8f_\text{s} \).}
		\label{Fig:res1p5}
	\end{center}
\end{figure}

Figure~\ref{Fig:res1p5} presents the FDTD results for space-time frequency mixing by a space-time-modulated Josephson metasurface operating at a modulation frequency of \( f_\text{s} =f_\text{0}= 1.5 \, \text{GHz} \). This configuration enables frequency generation of \(  2f_\text{s} \) through \( 8f_\text{s} \). 

In linear STM media, sideband generation is efficient when $\omega_\text{s}<<\omega_0$. As a result, linear STM media are not suitable for efficient frequency conversion, where high-ratio frequency conversion along with large signal frequency conversion is desired. In contrast, nonlinear STM media, particularly those involving Josephson junctions, enable efficient frequency conversion even when $\omega_\text{s}>\omega_0$. Here, we explain why efficient frequency conversion is achievable in nonlinear space-time-modulated media, unlike in their linear counterparts. In linear STM media, where the material parameter (e.g., permittivity) is modulated as $\epsilon(z,t)=\epsilon_\text{r}+\delta \cos(\omega_\text{s} t-\beta_\text{s} z)$,
	the modulation creates a periodic variation in the refractive index, which couples energy from the incident wave at $\omega_0$ to sidebands at frequencies 
	$\omega_0\pm n\omega_\text{s}$. The strength of these sidebands depends on the modulation depth $\delta$ and the phase-matching conditions. The following explains why sidebands are stronger when $\omega_\text{s}<<\omega_0$.
\begin{itemize}
		\item Dispersion effects: In linear media, dispersion cause the wavevector components of the sidebands to deviate from the phase-matching condition as $\omega_\text{s}$ increases. This mismatch reduces the efficiency of sideband generation at higher modulation frequencies.
		\item Phase-matching conditions: When $\omega_\text{s}<<\omega_0$, the spatial and temporal periodicity of the modulation aligns more closely with the wavelength of the incident wave. This phase-matching condition maximizes the overlap between the incident wave and the modulation, leading to efficient energy transfer to the sidebands.
		
		\item Effective interaction length: For low modulation frequencies ($\omega_\text{s}<<\omega_0$), the interaction length (over which the wave remains in phase with the modulation) is long, enhancing the energy coupling to the sidebands. As $\omega_\text{s}$ increases, the effective interaction length decreases, reducing the efficiency of sideband generation.
\end{itemize}

 The ability of these metasurfaces to efficiently convert frequencies at high modulation-to-signal frequency ratios ($\omega_\text{s} > \omega_0$) holds great promise for superconducting quantum technologies operating at millikelvin temperatures. These systems often require precise control over frequency conversion processes, and the nonlinear behavior allows for more flexible and efficient design of frequency converters, mixers, and modulators that operate at higher frequencies without the limitations imposed by linear systems. Fabrication of such metasurfaces may be accomplished through an array of Josephson junction field-effect transistors, which serve as highly tunable and low-loss components at cryogenic temperatures. These transistors leverage the quantum mechanical properties of Josephson junctions to achieve precise control over the modulation of electromagnetic waves at both microwave and millimeter-wave frequencies. By integrating these arrays into the metasurface architecture, it becomes possible to dynamically manipulate wave properties such as amplitude, phase, and frequency in real time. This approach not only ensures high efficiency and scalability but also opens avenues for advanced functionalities in quantum computing, nonreciprocal devices, and next-generation communication systems.

\bibliographystyle{IEEEtran}
\bibliography{Taravati_Reference}

\vfill

\end{document}